\documentclass[9pt,twocolumn,twoside]{osajnl}

\journal{ao} 

\setboolean{shortarticle}{false} 

\usepackage[colorinlistoftodos]{todonotes}
\usepackage[utf8]{inputenc}	
\usepackage{caption}

\usepackage{tabularx}
\usepackage{amsmath,amsfonts,amssymb}
\usepackage{graphicx}
\usepackage{float}

\title{Hybrid calibration procedure for fringe projection profilometry based on stereo-vision and polynomial fitting}

\author[1]{Raúl Vargas}
\author[1]{Andres G. Marrugo*}
\author[2]{Song Zhang}
\author[3]{Lenny A. Romero}

\affil[1]{Facultad de Ingeniería, Universidad Tecnológica de Bolívar, Cartagena, Colombia}
\affil[2]{School of Mechanical Engineering, Purdue University, West Lafayette, IN 47907, United States}
\affil[3]{Facultad de Ciencias Básicas, Universidad Tecnológica de Bolívar, Cartagena, Colombia}
\affil[*]{Corresponding author: agmarrugo@utb.edu.co}

\begin{abstract}
The key to accurate 3D shape measurement in Fringe Projection Profilometry (FPP) is the proper calibration of the measurement system. Current calibration techniques rely on phase-coordinate mapping (PCM) or back-projection stereo-vision (SV) methods. PCM methods are cumbersome to implement as they require precise positioning of the calibration target relative to the FPP system but produce highly accurate measurements within the calibration volume. SV methods generally do not achieve the same accuracy level. However, the calibration is more flexible in that the calibration target can be arbitrarily positioned. In this work, we propose a hybrid calibration method 
that leverages the SV calibration approach using a PCM method to achieve higher accuracy. The method has the flexibility of SV methods, is robust to lens distortions, and has a simple relation between the recovered phase and the metric coordinates. Experimental results show that the proposed Hybrid method outperforms the SV method in terms of accuracy and reconstruction time due to its low computational complexity.
\end{abstract}

\setboolean{displaycopyright}{true}

\begin{document}

\maketitle



\section{Introduction}
\label{sec:intro}

Accurate 3D shape measurement with Fringe Projection Profilometry (FPP) requires proper calibration~\cite{Zhang:2016tq,Zhang:2018jb,marrugo2018fourier}. Currently, most calibration procedures in FPP rely on phase-coordinate mapping (PCM) or stereo vision (SV) methods~\cite{cai2017phase,Li:2016hk,vargas2018camera}. 
PCM techniques relate experimental $XYZ$ metric coordinates to phase values using polynomial~\cite{vo2010flexible,huang2014new} or rational~\cite{huang2010least,zhou2014analysis,du2007three} fitting functions. However, they require the use of 2D or 3D targets with well-known dimensions, and positioning them at distances and orientations with high precision~\cite{Villa:2012jk}. Though accurate, these requirements lead to expensive calibration experimental setups~\cite{Zhang:2016tq}.


Alternatively, the extensively adopted SV method provides much more flexibility by calibrating the projector in a similar manner as a camera~\cite{zhang2006novel}. Hence, the FPP system can be modeled using the well-established camera calibration methods~\cite{hartley2003multiple}. The calibration process can be carried out using arbitrarily placed 2D targets, resulting in a more flexible calibration procedure~\cite{zhang2000flexible}. However, it is known that lens distortions introduce significant errors in the SV model 3D reconstruction~\cite{vargas2018camera,Vargas:2018vk}. Several strategies have been proposed to compensate for the projector and camera lens distortions to improve accuracy. Some consist of projecting additional patterns to obtain a direct camera-projector pixel correspondence~\cite{li2008accurate,moreno2012simple}. Although, they increase the acquisition time by requiring a large number of images or require costly correspondence search algorithms. Other methods consist of digitally pre-deforming the fringe pattern to obtain a distortion-free projected pattern~\cite{li2016lens,gonzalez2019accurate}. Nevertheless, the calibration procedure can become elaborate, which decreases the flexibility of the calibration procedure. 

In recent years, a different approach has emerged. Mainly, to combine PCM and SV models to obtain a flexible calibration procedure and a simpler calibration model~\cite{Chen:2016bi}. For instance, the SV calibration parameters were used to fit a polynomial PCM model, and such PCM model achieved the same accuracy as the SV model, but with higher computational efficiency~\cite{cai2017phase,cai2018ray}. Therefore, the question becomes, is it possible to use a PCM model to achieve higher accuracy?

To address the above question, we propose a four-stage hybrid calibration procedure. 1) we carry out a conventional SV calibration of the FPP system considering camera and projector lens distortions; 2) we obtain the pose of a flat board using the SV model 3) we compute pixel-wise error maps to ideal planes, 4) we obtain new pixel-wise relationship between ($\hat{X}$, $\hat{Y}$, and $\hat{Z}$) and phase values through pixel-wise polynomial fitting. We will demonstrate that our approach improves the overall accuracy while maintaining a low computational complexity.


\begin{figure*}[t]
    \centering
    \includegraphics[width=0.9\linewidth]{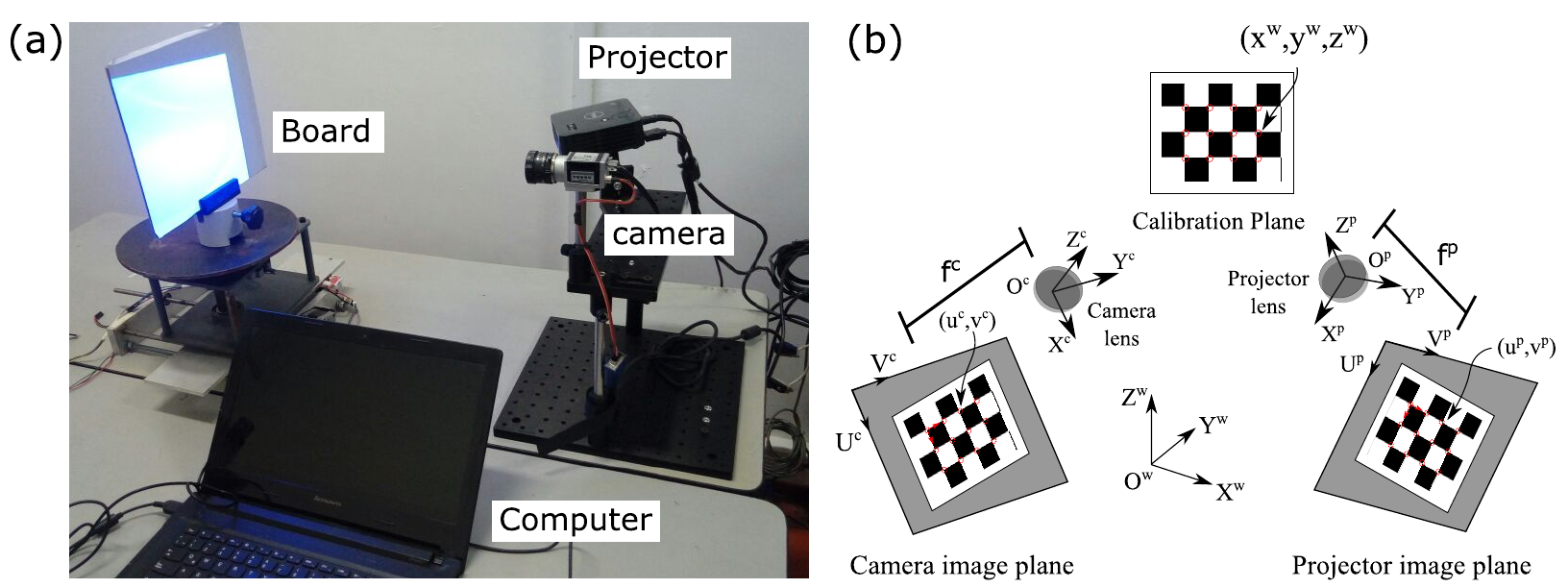}
    \caption{(a) Experimental Setup. (b) Stereo model diagram of a camera-projector system.}
    \label{fig:experimentalSetup-ModeloEstereo}
\end{figure*}

\section{Method and Materials}

\subsection{Experimental setup}

In Fig.~\ref{fig:experimentalSetup-ModeloEstereo}(a), we show the experimental setup, which consists of a monochromatic CMOS camera Basler Ace-1600gm with resolution of 1600x1200 at 60 fps with a focal length of 12 mm at F1.8 (Edmund Optics-58001), a DLP projector DELL M115HD with native resolution of 1280x800, a checkerboard for calibration with $8\times8$~mm squares, and a PC workstation. To synchronize the camera and DLP projector, we duplicate the VGA signal from the computer using a VGA splitter and connecting the vertical sync pulse to the hardware trigger pin of the camera.

\subsection{Stereo vision calibration}
\label{sec:SV}
In this work, we begin from the stereo calibration method proposed by Zhang and Huang~\cite{zhang2006novel} to develop a flexible and robust system calibration approach. The camera-projector system is considered as a binocular framework by regarding the projector as an inverse camera. In Fig.~\ref{fig:experimentalSetup-ModeloEstereo}(b), we show a schematic of the fringe projection system. Considering a point $(x^w, y^w, z^w)$ in the world coordinate system (WCS),  we find its corresponding coordinate in the camera or projector systems using the following equation,
%
\begin{equation}
\label{eq:proyeccionCam}
   s\begin{bmatrix}
   u \\ v\\1
   \end{bmatrix}= \begin{bmatrix}
  f_u&\gamma&c_u \\0&f_v &c_v\\0&0&1
   \end{bmatrix} \mathbf{M}_w
   \begin{bmatrix}
   \text{x}^w \\ \text{y}^w\\ \text{z}^w\\1
   \end{bmatrix},
\end{equation}
where $s$ is the scaling factor; $f_u$ and $f_v$ denote the effective focal lengths along $u$ and $v$ directions, respectively; $(c_u, c_v)$ is the coordinate of the principal point; $\gamma$ is the skew factor of images axes. The matrix $\mathbf{M}_w$ represents a rigid transformation from the world coordinate system~(WCS) to the camera coordinate system~(CCS) or the projector coordinate system~(PCS). Usually, the WCS matches with the CCS and the transformation WCS to PCS is defined as,
\begin{equation}
   ^p\mathbf{M}_w = [\mathbf{R}(\theta_s),\mathbf{t}_s],
\end{equation}
where $\textbf{R}$ is a rotation 3$\times$3 matrix, $\textbf{t}_s$ is a 3$\times$1 translation vector, and $\theta_s$ denotes a 3$\times$1 vector with the Euler angles.

The camera and projector distortion lens are modeled as~\cite{vo2011advanced}
\begin{align}
\begin{bmatrix}
u_d\\v_d
\end{bmatrix} =& (1+k_1r_n^2+k_2r_n^4+k_3r_n^6)\begin{bmatrix}
u_n\\v_n
\end{bmatrix}
+ \nonumber \\ &\begin{bmatrix}
    2p_1u_nv_n + p_2(r_n^2+2u_n^2) \\
2p_2u_nv_n + p_1(r_n^2+2v_n^2) \\
\end{bmatrix} \enspace,
\end{align}
with
\begin{align}
    r_n^2 &= u_n^2 + v_n^2 \enspace,
\end{align}
where $[k_1, k_2, k_3]$ and $[p_1, p_2]$ are the radial and tangential distortion coefficients, respectively; $[u_d, v_d]^t$ and $[u_n, v_n]^t$ refer to normalized coordinate before and after the distortion correction.

\begin{figure*}[t]
    \centering
    \includegraphics[width=0.72\linewidth]{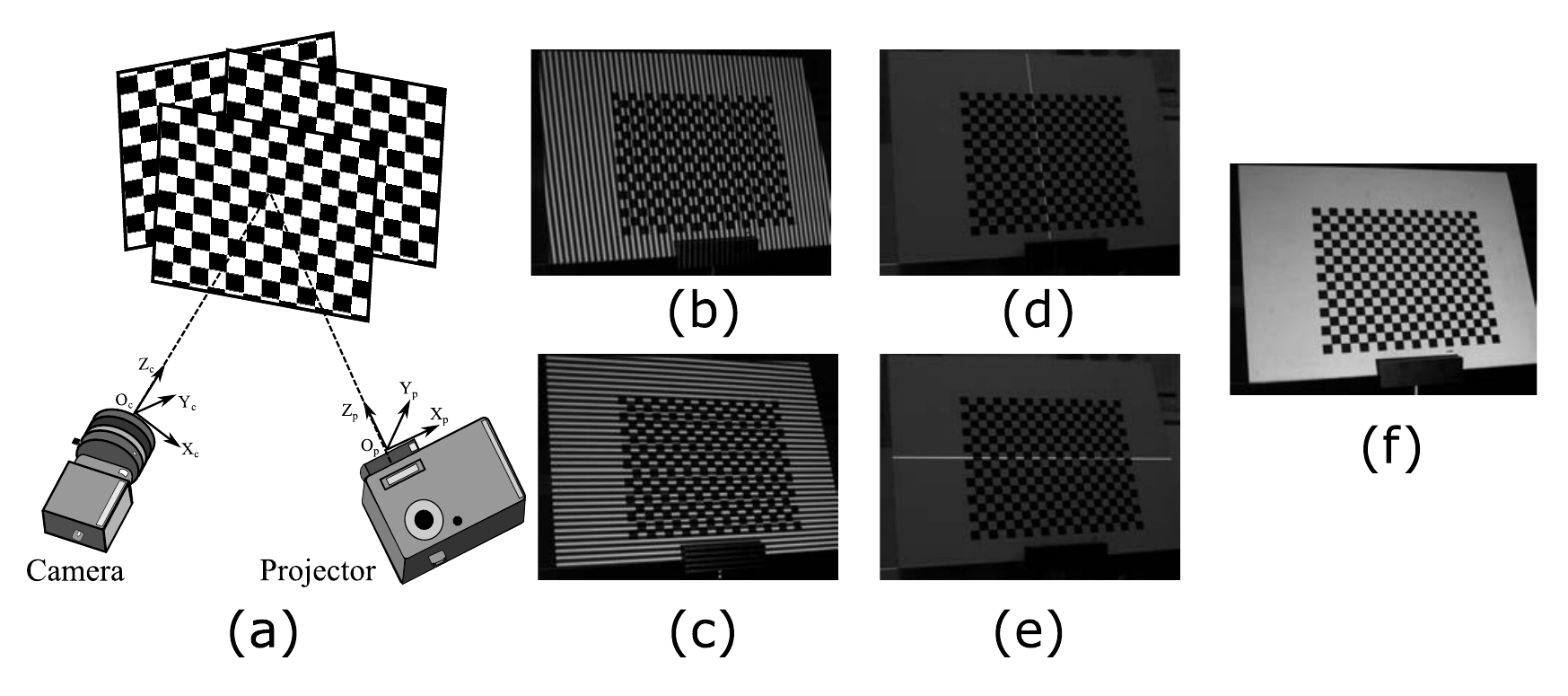}
    \caption{The FPP system is initially calibrated using the conventional (a) Stereo calibration procedure with a B\&W checkerboard. It requires capturing (b)-(f) vertical and horizontal fringe patterns, along with centerline images for absolute phase retrieval, and a texture image.}
    \label{fig:SVprocedure}
\end{figure*}


To obtain the stereo parameters that describe our system, we use a standard stereo calibration procedure using a black and white (B\&W) checkerboard. We placed the board at 15 different distances and orientations from the system, as shown in Fig~\ref{fig:SVprocedure}(a). In each position, we captured the images shown in Fig.~\ref{fig:SVprocedure}(b)-(f). Using the vertical (Fig.~\ref{fig:SVprocedure}(b)) and horizontal (Fig.~\ref{fig:SVprocedure}(c)) fringe images, we recovered the discontinuous phase employing the phase-shifting algorithm with 8 images. Afterward, we apply a phase unwrapping algorithm using a centerline image (Fig.~\ref{fig:SVprocedure}(d) and Fig.~\ref{fig:SVprocedure}(e)) to obtain the absolute  phase maps in the horizontal ($\phi_h$) and vertical ($\phi_v$) directions\cite{zhang2006novel}.  Additionally, we capture a texture image of the checkerboard projecting white light on it (Fig.~\ref{fig:SVprocedure}(f)).

We calculated the corners $(\hat{u}^c, \hat{u}^c)$ with subpixel precision using texture images, and corner coordinates in the projector image plane $(\hat{u}^p, \hat{u}^p)$ were calculated using phase values in each detected corner. To avoid phase errors in the checkerboard corners, we used only the phase from the white squares (having morphologically eroded boundaries to avoid black-to-white transition), and we interpolate the phase values using a 5th order 2D polynomial function to all pixels following a similar procedure as described in Ref.~\cite{Zhang:2011ea}.

Having determined the coordinates $(\hat{u}^c, \hat{u}^c)$ and $(\hat{u}^p, \hat{u}^p)$, we used the camera calibration toolbox proposed by Bouguet~\cite{bouguet2008camera} to obtain the intrinsic and extrinsic stereo parameters. The obtained reprojection errors for the camera and the projector are 0.154 and 0.099, respectively, which are quite small.
In Table~\ref{Tab:SVParameters}, we show the stereo parameters obtained for our system.


\begin{table}[t]
\centering
\caption{System stereo parameters of SV calibration. }
\begin{tabular}{ccc}
\hline
Parameter & Camera & Projector \\
\hline
$[c_u,c_v]$ [pix]& 794.59, 594.32    & 632.42, 799.67    \\
$[f_u,f_v]$ [pix]& 2698.42, 2701.55 & 1949.18, 1953.15  \\
$\gamma$&  0.84 &  0.83 \\
$\mathbf{p}_1 $&  0.0011   &   0.0004    \\
$\mathbf{p}_2 $&  0.0006    &   0.0002    \\
$\mathbf{k}_1$& -0.12  &  0.002   \\
$\mathbf{k}_2$& -0.52   & -0.077  \\
$\mathbf{k}_3$& 5.08 &   -0.033 \\
$\theta_s$ [$^\circ$] & \multicolumn{2}{c}{-0.42, 19.08, 0.36 }\\
$\mathbf{t}_s$ [mm] & \multicolumn{2}{c}{-163.71, -116.73, 135.41}\\
\hline
\end{tabular}
  \label{Tab:SVParameters}
\end{table}

\begin{figure}[t]
    \centering \includegraphics[width=0.90\linewidth]{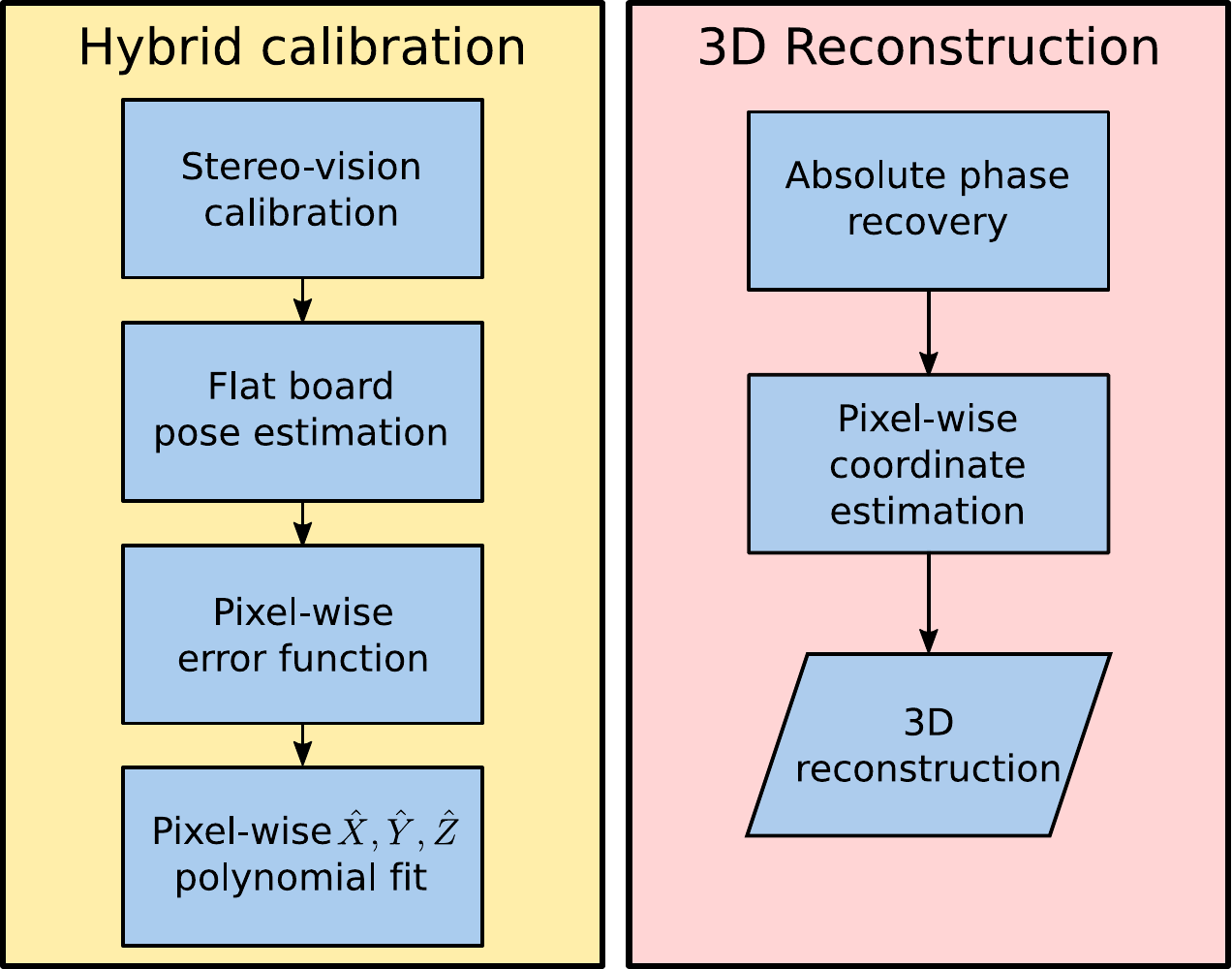}
    \caption{The hybrid calibration procedure, and the resulting 3D reconstruction method.}
    \label{fig:block-diagram}
\end{figure}

\subsection{Hybrid calibration}

In Fig.~\ref{fig:block-diagram}, we show a block diagram of the proposed hybrid calibration procedure. We begin from the standard SV calibration approach. We then use it to obtain the pose of a flat white board at different $n$ positions. Suppose that any point ($X, Y, Z$) of the 3D reconstruction of each pose $n$, will have small residual errors~$(\delta_x, \delta_y, \delta_z)$ relative to the true 3D surface. We assume that the best guess to the true surface is an ideal plane estimated via least-squares. Accordingly, the residual errors are modeled as the perpendicular distances to that plane. For each reconstructed 3D point along the line-of-sight of a given camera pixel $(u_c, v_c)$ we calculate the errors relative to each ideal plane. This correction does not generally result in a straight line, but in a smooth curve in space. Finally, we use a pixel-wise third-order polynomial regression model to relate the absolute phase values to the corrected ($\hat{X}, \hat{Y}, \hat{Z}$)  metric coordinates.

\subsubsection{Flat board pose estimation}
The mapping procedure that we propose consists of positioning a flat white board within a predefined calibration volume and move it through that volume, as shown in Fig.~\ref{fig:regressionProcedure}(a). We propose to use a white board since it allows us to have a phase map without the presence of errors caused by intensity variations. Although the plane can be placed arbitrarily throughout the calibration volume, we decided to manually move the plane in 30 positions using homogeneous displacements. This operation guarantees a correct sampling of the region of interest since fitting a third-order polynomial requires a minimum of four experimental phase-coordinate relations. In each position $n$, the planes are reconstructed using the SV model, obtaining the respective $X$, $Y$, and $Z$ coordinates for each camera pixel $(u_c, v_c)$. 

\subsubsection{Pixel-wise error function}
\begin{figure}[t]
    \centering
    \includegraphics[width=0.85\linewidth]{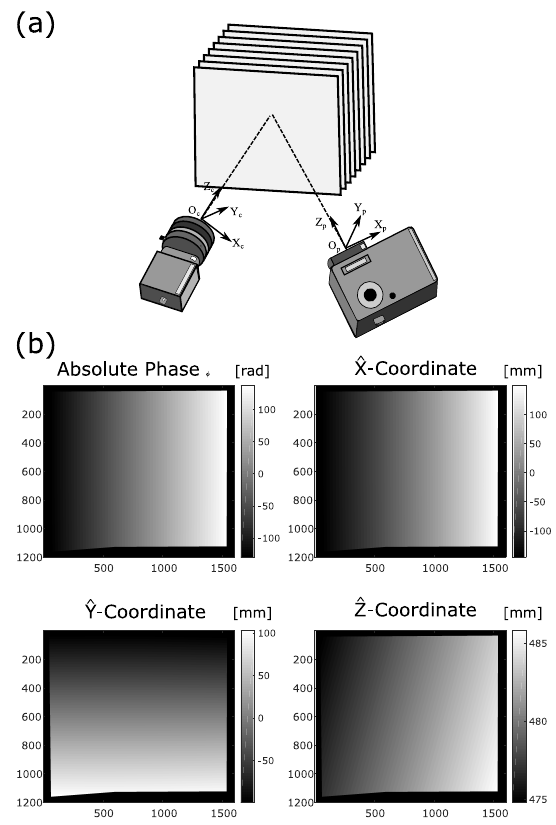}
    \caption{The (a) Hybrid calibration procedure consists in manually displacing a flat white board along a predefined calibration volume. The (b) recovered absolute phase map $\phi$ and metric coordinates ($\hat{X},\hat{Y},\hat{Z}$) of an ideal plane estimated for a position of the flat board are used for the input for the hybrid calibration. }
    \label{fig:regressionProcedure}
\end{figure}

Nevertheless, the SV-reconstructed coordinates have residual errors. To reduce these errors, we fit each reconstructed plane to an ideal plane by least-squares. Then, each $X$, $Y$, and $Z$ coordinate of the reconstructed plane is corrected as follows,
\begin{align}
    \hat{X} &= X - \delta_x \enspace,  \\
    \hat{Y} &= Y - \delta_y \enspace,  \\
    \hat{Z} &= Z - \delta_z \enspace.
    \label{eq:new_coords}
\end{align}
where $\hat{X}$, $\hat{Y}$, and $\hat{Z}$ represent the corrected coordinates for a camera pixel ($u_c, v_c$),  $(\delta_x, \delta_y, \delta_z)$ are the estimated residual errors from the perpendicular distance between the ideal plane and each point ($X$, $Y$, $Z$). In Fig.~\ref{fig:regressionProcedure}(b), we show the phase map using vertical fringes and the corrected coordinates for the plane at position $n = 15$.

\subsubsection{Pixel-wise polynomial fit}

We use a pixel-wise regression model to relate the absolute phase values to the metric coordinates of the FPP system for each pixel $(u_c, v_c)$,
\begin{align}
    \hat{X} &=  a_3\phi^3+ a_2\phi^2+ a_1\phi+ a_0 \enspace, \label{eq:reg1} \\
    \hat{Y} &=  b_3\phi^3+ b_2\phi^2+ b_1\phi+ b_0 \enspace, \label{eq:reg2}\\
    \hat{Z} &=  c_3\phi^3+ c_2\phi^2+ c_1\phi+ c_0 \label{eq:reg3}\enspace,
\end{align}
where $\phi $ is the absolute phase value for a camera pixel ($u_c, v_c$); $\hat{X}$, $\hat{Y}$, and $\hat{Z}$ are the corrected metric coordinates in the camera coordinate system; and $a_{0-3} $, $b_{0-3} $ and $c_{0-3} $ are the coefficients of the adjustment polynomials $\phi-X$, $ \phi-Y$ and $\phi-Z$, respectively. We have omitted the pixels $(u_c, v_c)$ in the notation to avoid making the equations more complicated.

\begin{figure}[t]
    \centering
    \includegraphics[width=\linewidth]{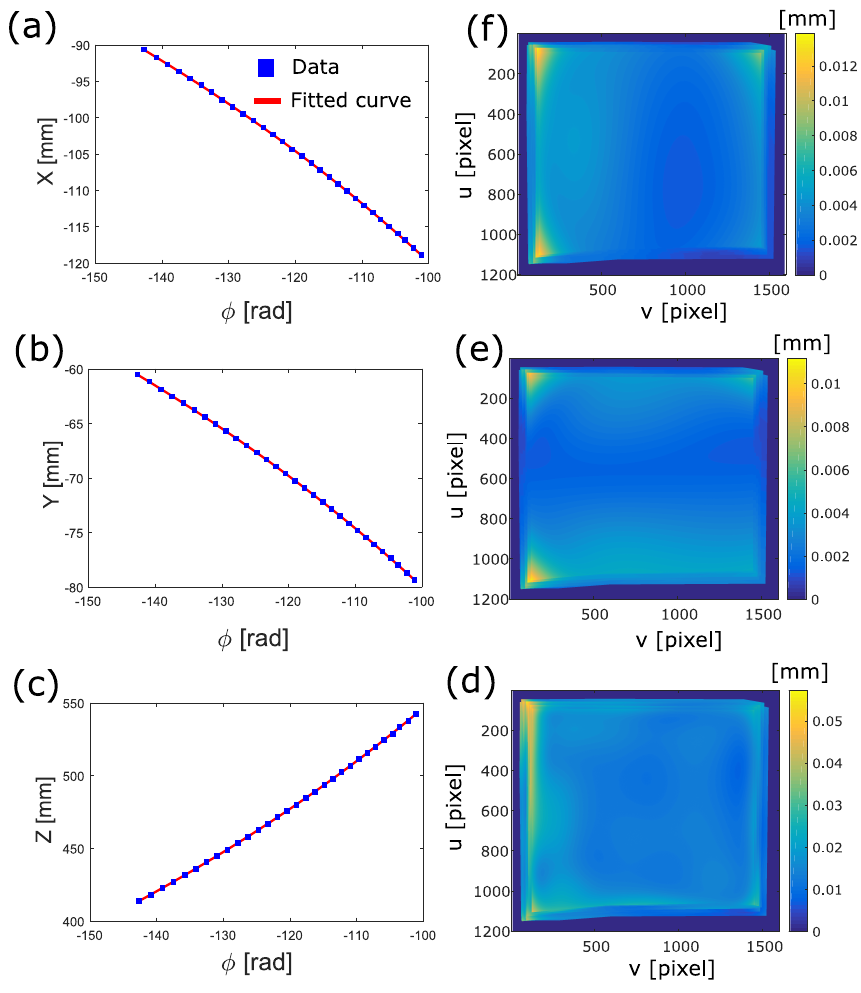}
    \caption{A third-degree polynomial function describes sufficiently well the phase $\phi$ to ideal (a) $\hat{X}$, (b) $\hat{Y}$ and (c) $\hat{Z}$ coordinates of the image pixel (200, 200). The RMS error maps of (d)~$\hat{X}$, (e)~$\hat{Y}$ and (f)~$\hat{Z}$-calibration show a small error throughout most of the field of view.}
    \label{fig:RegressionResults}
\end{figure}

Figures~\ref{fig:RegressionResults}(a),~(b)~and~(c) show the experimental data and the adjustment polynomial between the phase values and the $\hat{X}$, $\hat{Y}$ and $\hat{Z}$ coordinates for the pixel $(200,200)$ of the camera. Note that the third-order polynomial allows obtaining a correct representation of the data. 
We calculate an RMS error between the $\hat{X}$, $\hat{Y}$ and $\hat{Z}$ corrected experimental data and the fitted polynomials for each pixel for all poses of the white flat board. In Figures~\ref{fig:RegressionResults}(d)-(f) we show the RMS errors corresponding to $\phi-\hat{X}$, $\phi-\hat{Y}$, and $\phi-\hat{Z}$, respectively.
The RMS errors of the polynomial fittings have a maximum of 0.060~mm for the $\hat{Z}$ coordinate, while for the $\hat{X}$- and $\hat{Y}$- coordinates, the RMS errors are less than 0.015~mm. Therefore, the third-order polynomial regression model correctly describes our system within the calibrated volume, which is approximately 270~mm $\times$ 200~mm $\times$ 130~mm.

\subsubsection{3D reconstruction}

As shown in Fig.~\ref{fig:block-diagram}, the 3D reconstruction under the proposed hybrid calibration procedure consists of two stages. First, recovering the absolute phase per camera pixel from the projected vertical fringe patterns. Second, evaluating the polynomials from equations \eqref{eq:reg1}-\eqref{eq:reg3} to obtain the accuracy-improved 3D reconstruction.

\section{Experimental Results}

\subsection{Experiment 1: Reconstruction of a flat board}

\begin{figure}[t]
    \centering \includegraphics[width=0.95\linewidth]{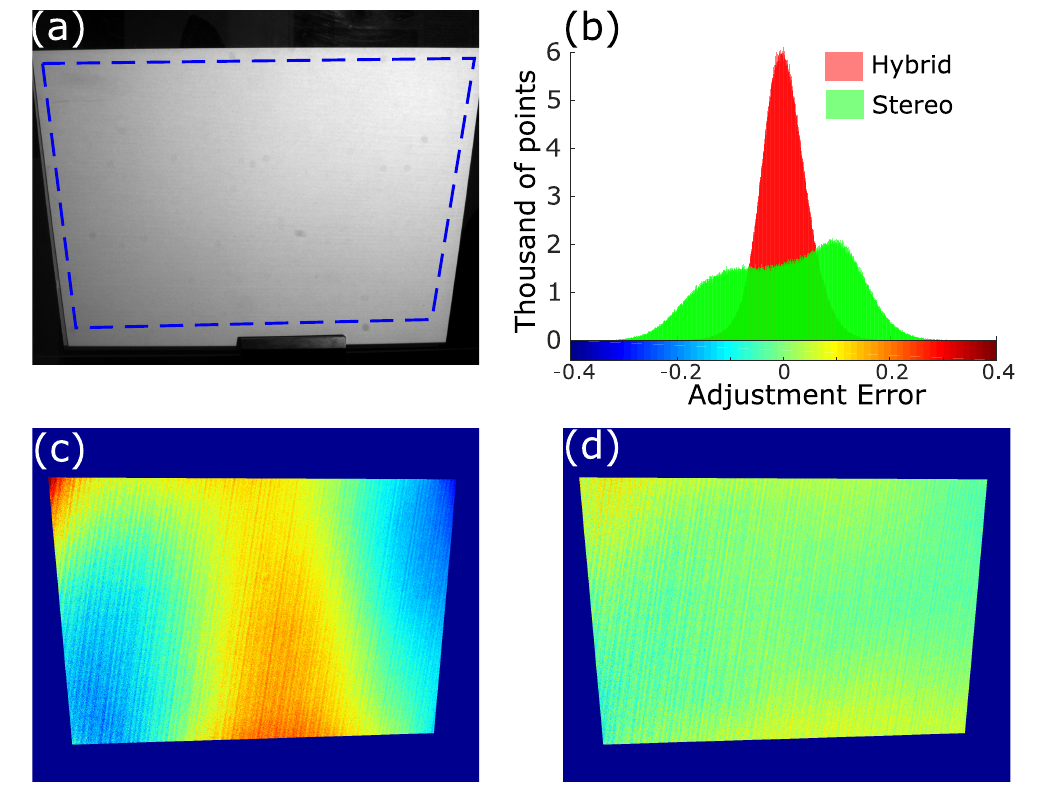}
    \caption{\textit{Experiment 1} results. (a) Flat board. (b) Error histograms in the reconstruction of a flat board using SV, and hybrid calibration models. (c) and (d) adjustment error maps of a plane reconstructed by SV and hybrid model, respectively. The hybrid model significantly improves the accuracy of the 3D reconstruction.}
    \label{fig:exp1}
\end{figure}

Here, we compare the accuracy of the SV and Hybrid models.  To do this, we reconstructed a flat white board of 230~mm~$\times$~175~mm positioned within the calibration volume of the hybrid model, and we inclined it relative to the $Z$ axis of the camera, as shown in Fig.~\ref{fig:exp1}(a). This board was reconstructed using the two calibration models, and an adjustment to an ideal plane was carried out by least-squares fitting. In Fig.~\ref{fig:exp1}(b), we show the histograms of the plane adjustment error corresponding to each calibration model. Note that the hybrid calibration model improves the accuracy of the reconstruction of the plane, reducing the dispersion of the error distribution and obtaining an RMS error of 0.044~mm, which is lower than the RMS error of 0.118~mm obtained with the SV model. Therefore, although the hybrid model is derived from the SV model, it provides greater precision in the reconstruction by compensating the residual errors. In the figures~\ref{fig:exp1}(c)~and~\ref{fig:exp1}(d), we show the error maps obtained with the SV and the hybrid model, respectively. In Fig.~\ref{fig:exp1}(c), we can see that the residual errors in the SV model vary through the field of view in a range of $[-0.2, 0.2]$~mm, while the hybrid model has a more homogeneous error distribution with most errors in the range of $[-0.1, 0.1]$~mm.

\subsection{Experiment 2: Reconstruction of a cylinder}
\begin{figure}[t]
    \centering \includegraphics[width=0.95\linewidth]{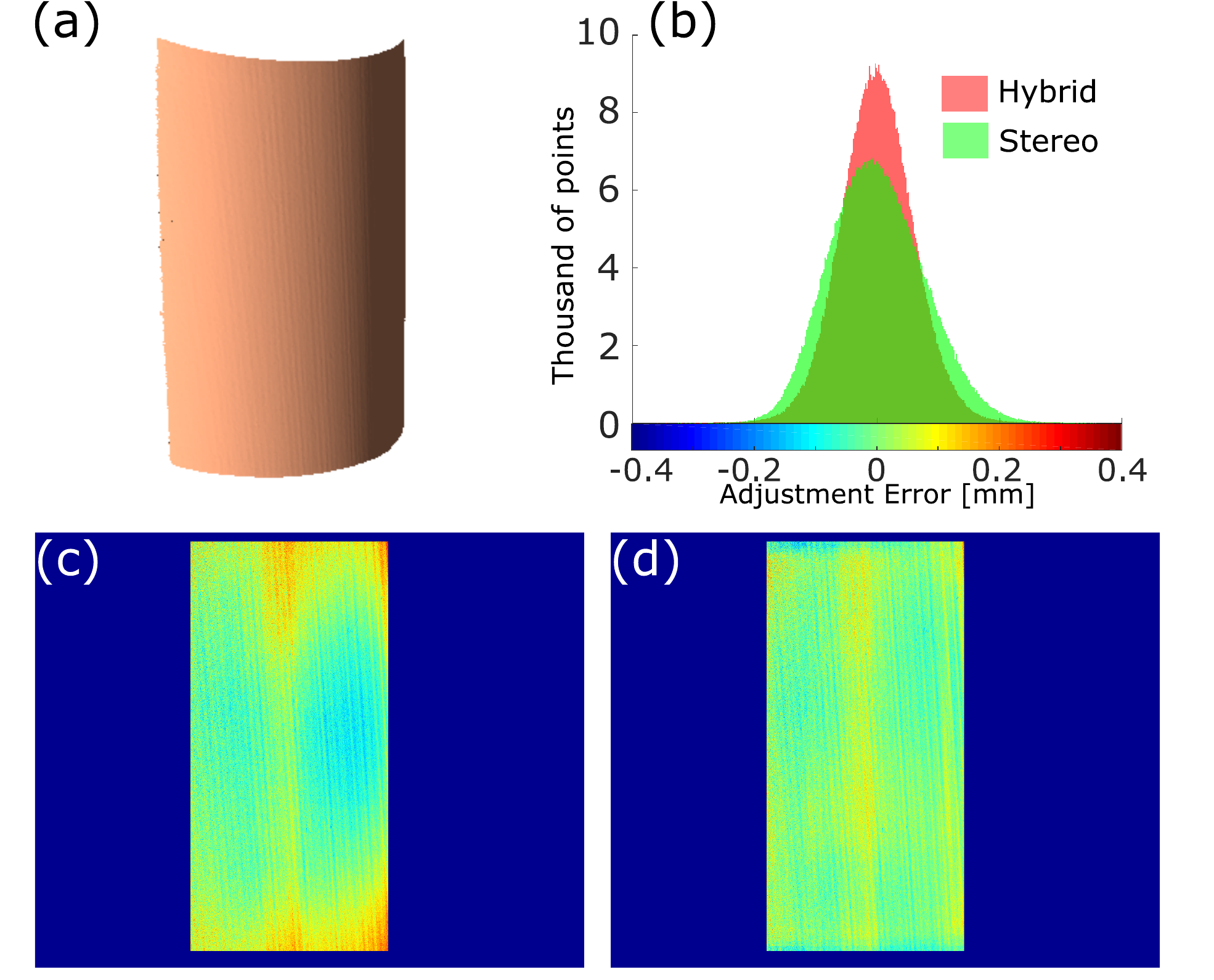}
    \caption{\textit{Experiment 2} results. (a) 3D reconstruction shape of a cylinder using the hybrid model. (b) Histogram of Adjustment error of a cylinder reconstructed using the hybrid and SV models. (c) and (d) Error maps of the cylinder reconstructed by SV model and hybrid model, respectively. The accuracy improvement is also displayed in the 3D reconstruction of a non-flat object.}
    \label{fig:exp2}
\end{figure}

In this experiment, we evaluate the performance of the proposed calibration model to perform the reconstruction of a non-flat object, a cylinder. The tested object is a 168mm diameter and 215mm height Polyvinyl chloride (PVC) pipe. A section of the pipe was reconstructed using the SV and Hybrid models. Fig.~\ref{fig:exp2}(a) shows the 3D shape of the cylindrical cap reconstructed using the Hybrid model. To assess the shape measurement capability of each method, we adjust each reconstruction to an ideal cylinder by least-squares fitting. The histograms of adjustment errors obtained for each reconstruction are shown in Fig~\ref{fig:exp2}(b). Note that the hybrid model allows obtaining a correct cylindrical representation of the reconstruction since the errors are more concentrated around zero, while the histogram of the SV model has a higher dispersion. The RMS adjustment errors obtained with the SV and hybrid models were 0.076 and 0.059 mm, respectively. Figures~\ref{fig:exp2}(c)~and~(d) show the adjustment error maps for the SV and hybrid models, respectively. Note that in the SV model, the largest residual errors are located in the upper and lower part of the cylinder; meanwhile, in the error map of the hybrid reconstruction, the errors are uniformly distributed throughout the image.

\subsection{Experiment 3: Complex surface reconstruction}
\begin{figure}[t]
    \centering \includegraphics[width=0.95\linewidth]{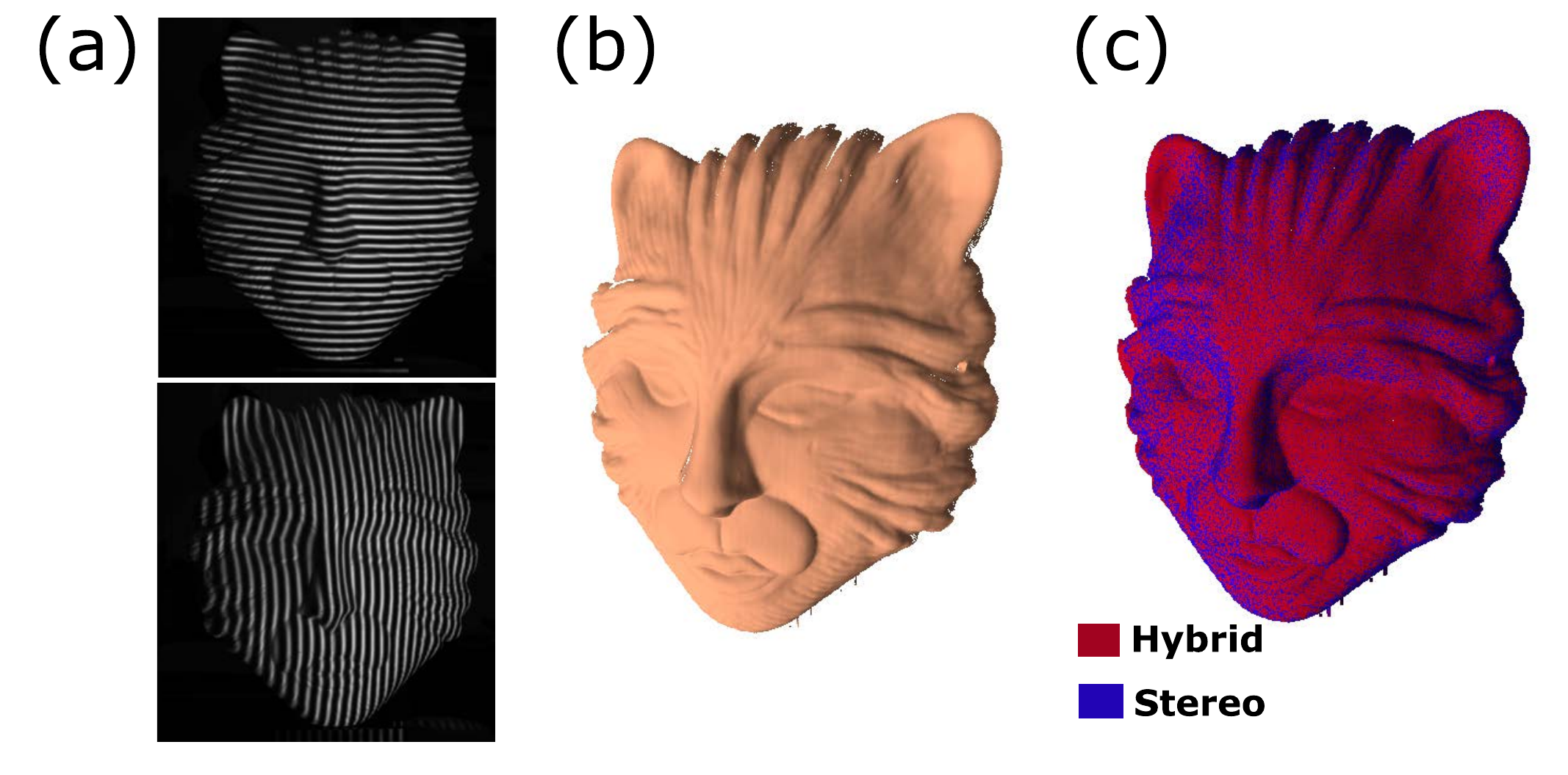}
    \caption{\textit{Experiment 3} results. (a) Captured fringe images of a ceramic mask. (b) 3D reconstruction of the mask. (c) Superposition of the 3D reconstructions using the hybrid and SV models. }
    \label{fig:exp4}
\end{figure}

In this experiment, we performed the reconstruction of a complex ceramic mask of an approximate size of 160mm$\times$190mm. To do this, we projected vertical and horizontal fringe patterns onto the surface of the object, as shown in Fig.~\ref{fig:exp4}(a). It was necessary to project the horizontal strip patterns to compensate the distortions of the projector in the stereo model, while the hybrid model only requires the projection of vertical fringes. In Fig.~\ref{fig:exp4}(b), we show the reconstruction of the ceramic mask using the hybrid model. Fig.~\ref{fig:exp4}(c) shows the superposition of the reconstructions of the hybrid and SV models. The RMS distance between the two reconstructions is 0.075~mm. The reconstruction times of the object (683047 points) were 1.04s and 0.14s for the SV models and the hybrid model, respectively.

\subsection{Experiment 4: Execution time assessment}

\begin{figure}[t]
    \centering
    \includegraphics[width=1\linewidth]{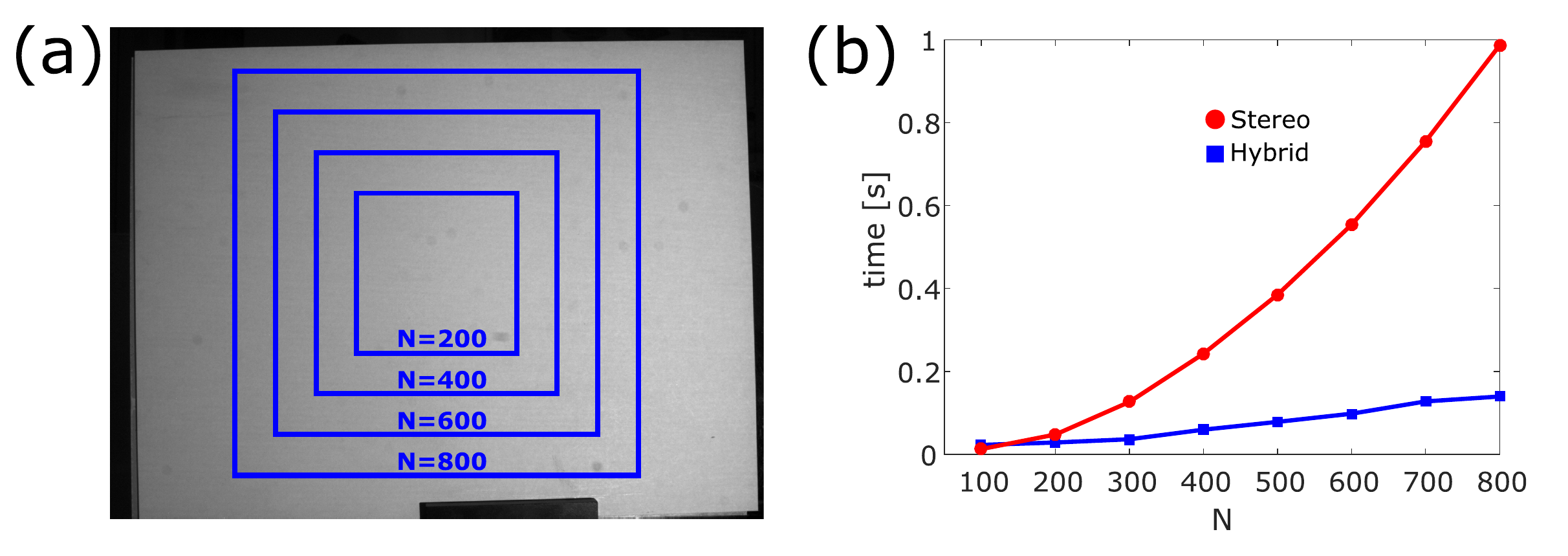}
    \caption{\textit{Experiment 4} results. (a) Reconstructed NxN size windows of an object for reconstruction time analysis. (b) reconstruction time curves of the SV and hybrid models using square reconstruction windows with NxN pixels. }
    \label{fig:exp5}
\end{figure}

Finally, we analyzed the execution times required for the hybrid model and the SV model to reconstruct objects of different sizes. To simulate this, we reconstructed several sections of a study object with different numbers of pixels. The object of reconstruction is the flat board used in the calibration process, and the reconstruction sections were chosen using square windows with a size of $NxN$ pixels, as shown in Fig.~\ref{fig:exp5}(a). 

In Fig.~\ref{fig:exp5}(b), we show the time required to perform the reconstructions of eight different window sizes using the Hybrid and SV models. Note that the hybrid model performs the 3D reconstructions with low computation time compared to the stereo model. This difference is because the SV model demands significant time in the compensation of distortions of the projector, which is carried out with an iterative process. Whereas in the hybrid model, these distortions are implicitly compensated in the polynomial model coefficients for each pixel, and no iterative computation is required.

\section{Conclusion}

Improving the measurement accuracy of a standard fringe projection profilometry system typically requires elaborate calibration procedures or sophisticated iterative methods to reduce 3D reconstruction error. Here, we have proposed a calibration method that leverages the stereo calibration approach for fringe projection profilometry and improves the measurement accuracy. The hybrid method uses the reconstructions of a flat white board as input to obtain pixel-wise polynomials for converting phase to metric coordinates with higher accuracy than the underlying stereo model. The accuracy improvement was shown through several experiments. Moreover, the low computational complexity of the proposed method reduces the execution time for 3D reconstruction significantly.

\section*{Funding}
Colciencias (project 538871552485), and Universidad Tecnológica de Bolivar (projects C2018P005 and C2018P018).

\section*{Disclosures}
The authors declare no conflicts of interest.

\section*{Acknowledgement}
R.~Vargas thanks Universidad Tecnológica de Bolívar (UTB) for a post-graduate scholarship. L.A. Romero and A.G. Marrugo thank UTB for a Research Leave Fellowship. A.G. Marrugo acknowledges support from the Fulbright Commission in Colombia and the Colombian Ministry of Education within the framework of the Fulbright Visiting Scholar Program, Cohort 2019-2020. Parts of this work were presented at the 2019 Iberoamerican Optics Meeting (RIAO) in Cancun, Mexico, and at the 2019 SPIE Optical Metrology conference in Munich, Germany~\cite{vargas2019flexible}.

\bibliography{report.bib}

\end{document}